\documentclass[preprint,aps,pra,showpacs,floatfix]{revtex4-1}
\usepackage[utf8]{inputenc}
\usepackage[OT1]{fontenc}
\usepackage{amsmath}
\usepackage{amsfonts}
\usepackage{amssymb}
\usepackage{bm}
\usepackage{color}
\usepackage{graphicx}
\usepackage[left=2cm,right=2cm,top=2cm,bottom=2cm]{geometry}
\usepackage{longtable}
\allowdisplaybreaks
\usepackage{booktabs}
\usepackage{longtable}
\usepackage{siunitx}
\usepackage{multirow}

\newcommand{\balpha}{\bm{\alpha}}

\newcommand{\br}{\bm{r}}

\newcommand{\be}{\begin{eqnarray}}
\newcommand{\ee}{\end{eqnarray}}

\usepackage{ulem}
\usepackage{color}

\begin{document}

\title{Binding energies of the $1s^2\,2s\,2p\, ^3P_{0,2}$ states of berylliumlike xenon}

\author{A.~V.~Malyshev$^{1,2}$, D.~A.~Glazov$^{1}$, Yu.~S.~Kozhedub$^{1}$, M.~Yu.~Kaygorodov$^{1}$, I.~I.~Tupitsyn$^{1,3}$, V.~M.~Shabaev$^{1}$, and G.~Plunien$^{4}$}

\affiliation{
$^1$ Department of Physics, St. Petersburg State University, Universitetskaya 7/9, 199034 St. Petersburg, Russia \\
$^2$ ITMO University, Kronverkskiy 49, 197101 St.~Petersburg, Russia \\
$^3$ Center for Advanced Studies, Peter the Great St. Petersburg Polytechnic University,
Polytekhnicheskaja 29, 195251 St.~Petersburg, Russia \\
$^4$ Institut f\"ur Theoretische Physik, Technische Universit\"at Dresden,
Mommsenstra{\ss}e 13, D-01062 Dresden, Germany 
}

\begin{abstract}
Binding energies of the $^3P_0$ and $^3P_2$ levels of the $1s^2\,2s\,2p$ electron configuration in berylliumlike xenon are rigorously evaluated using \textit{ab initio} QED approach. All relevant one- and many-electron QED contributions are accounted for up to the second order of the perturbation theory. The interelectronic-interaction effects of the third and higher orders are considered within the Breit approximation. Nuclear recoil effect is taken into account as well. In addition, we study all possible levels of the configuration $1s^2\,2s\,2p$, namely $^1P_1$ and $^3P_{0,1,2}$, by means of the configuration-interaction Dirac-Fock-Sturm method in berylliumlike neon, iron, and xenon. In this case the QED effects are treated approximately within the model QED approach. The obtained theoretical predictions are compared with the results of previous relativistic calculations and high-precision measurements. 
\end{abstract}

\maketitle

\section{Introduction}

High-precision measurements of the Lamb shift in H-like~\cite{Stoehlker:2000:3109,Gumberidze:2005:223001} and Li-like~\cite{Schweppe:1991:1434,Brandau:2003:073202,Beiersdorfer:2005:233003} uranium which provided tests of bound-state quantum electrodynamics (QED) with an unprecedented accuracy have triggered a new tide of interest in spectroscopy of highly charged ions both from experimental and theoretical sides. Perfect agreement of \textit{ab initio} QED calculations of the $2p_{3/2}-2p_{1/2}$ fine-structure splitting in B-like argon~\cite{Artemyev:2007:173004,Artemyev:2013:032518,Malyshev:2017:103} with the corresponding measurements~\cite{Draganic:2003:183001,Mackel:2011:143002} illustrates the extremely fruitful joint effort of experimentalists and theorists in this direction. On the other hand, one should mention also the disagreement between the experiment and the most accurate to-date evaluation of the transition energies in He-like ions~\cite{Artemyev:2005:062104}. The discrepancy has been claimed on the grounds of the high-precision measurements with heliumlike titanium~\cite{Chantler:2012:153001,Chantler:2014:123037}. This problem has motivated a series of new experiments to measure the X-ray transition energies in middle-$Z$ He-like ions~\cite{Rudolph:2013:103002,Kubicek:2014:032508,Beiersdorfer:2015:032514,Epp:2015:020502_R,Machado:HeBe:preprint}. Recently, we have studied the contribution of the nuclear recoil effect to the energies of the ground and low-lying excited states in heliumlike ions \cite{Malyshev:2017:recoil:preprint}. It was shown that the nuclear recoil can not be responsible for this discrepancy. The work has to be continued in order to finally  clarify the situation.

In the present work we study the binding energies of the singly excited states of Be-like ions which are also of experimental interest~\cite{Beiersdorfer:1998:1944,Trabert:2003:042501,Draganic:2003:183001,Feili:2005:48,Schippers:2012:012513,Bernhardt:2015:144008}. There are many relativistic calculations of the energy levels in berylliumlike ions~\cite{Safronova:1996:4036,Chen:1997:166,Safronova:2000:1213,Majumder:2000:042508,Gu:2005:267,Ho:2006:022510,Cheng:2008:052504,Sampaio:2013:014015,Yerokhin:2014:022509,Wang:2015:16,Li:2017:720}. These calculations include many-electron QED effects at best within some one-electron approximation. In Ref.~\cite{Bernhardt:2015:144008}, where the intra-L-shell transitions in Be-like xenon have been investigated employing the process of dielectronic recombination, it was noted that different existing theoretical approaches show significant scatter of the results. A more rigorous treatment of the QED effects should improve the theoretical accuracy~\cite{Yerokhin:2014:022509}. In our previous works~\cite{Malyshev:2014:062517,Malyshev:2015:012514} we have performed \textit{ab initio} QED calculations of the ground-state energies of Be-like ions, i.e., the energies for a system with the closed electron shells only. The developed method merges the rigorous QED calculations within the first and second orders of the perturbation theory including one- and many-electron QED contributions with the evaluation of the third- and higher-order electron-interaction effects in the framework of the Breit approximation. Later, the method was applied to the calculations of the ground-state ionization energies of B-like ions~\cite{Malyshev:2017:103,Malyshev:2017:022512} which represent the system with one valence electron. In the present work we have extended the approach developed for the closed shells and the one electron over the closed shells to the case of a single level with two valence electrons over the closed shell. With this method we have performed high-precision QED calculations of the $^3P_0$ and $^3P_2$ levels of the $1s^2\,2s\,2p$ electron configuration in Be-like xenon. The development of \textit{ab initio} QED approach for evaluation of the quasidegenerate levels $^1P_1$ and $^3P_1$ of the same electron configuration is in progress now. 

In order to evaluate the binding energies of the excited states of Be-like ions, we employ also the alternative simplified approach in addition to \textit{ab initio} method. Within the alternative numerical scheme the correlation effects are treated by the relativistic configuration-interaction Dirac-Fock-Sturm (CI-DFS) method~\cite{Bratzev:1977:2655,Tupitsyn:2003:022511}, while the QED corrections are considered using the model Lamb-shift operator \cite{Shabaev:2013:012513,Shabaev:2013:175}. The approximate approach is similar to the procedure which was used for calculations of the core-excited states in lithiumlike ions with $Z\leqslant 36$ in Refs.~\cite{Yerokhin:2012:042507,Yerokhin:2017:042505} and for calculations of the energy levels in berylliumlike iron ($Z=26$) in Ref.~\cite{Yerokhin:2014:022509}. Employing the simplified approach we have evaluated the $^1P_1$ and $^3P_{0,1,2}$ energy levels of the configuration $1s^2\,2s\,2p$ in Be-like neon, iron, and xenon.


The relativistic units ($\hbar=c=1$) are used throughout the paper. The CODATA 2014 recommended values of the fundamental constants \cite{Mohr:CODATA2014} are used:
$\alpha^{-1}=137.035999139(31)$ and $mc^2=0.5109989461(31)$~MeV.

         
\section{Theoretical approach}

The natural zeroth-order approximation to construct QED perturbation series for highly charged ions is provided by the $jj$-coupling scheme, in which unperturbed wave functions are constructed from the solutions of the one-electron Dirac equation
\begin{equation}
\left[-i \balpha \cdot \nabla + \beta m + V ( \br )\right] \psi_n ( \br ) = \varepsilon_n \psi_n ( \br ) \, .
\label{DirEq}
\end{equation}
Choosing the potential $V(\br)$ in Eq.~(\ref{DirEq}) to be the potential of the nucleus $V_{\rm nucl}(\br)$ leads to quantum electrodynamics in the Furry picture \cite{Furry:1951:115}. This choice is not the unique possible way to determine the initial approximation. Indeed, one can take the potential $V(\br)$ to be a sum of the nuclear potential and some local screening potential modeling the interelectronic-interaction effects
\begin{equation}
V( \br ) \rightarrow V_{\rm{eff}}( \br ) = V_{\rm{nucl}}( \br ) + V_{\rm{scr}}( \br ).
\label{EffPot}
\end{equation}
This choice of the zeroth order approximation corresponds to the so called extended Furry picture. Rearrangement of the perturbation series induced by the inclusion of the proper screening potential generally allows for the acceleration of the convergence of these series, since it involves the higher-order corrections partially. Obviously, that the counterterm $\delta V(\br) = -V_{\rm scr}(\br)$ has to be calculated perturbatively in this case in order not to consider the screening effects twice.

In the present paper we have performed \textit{ab initio} QED calculations of the $^3P_{0}$ and $^3P_{2}$ levels of the $1s^2\,2s\,2p$ electron configuration both with the Coulomb potential of the nucleus in Eq.~(\ref{DirEq}) (the standard Furry picture) and with the local Dirac-Fock (LDF) and core-Hartree (CH) potentials included to the zeroth-order approximation (the extended Furry picture). We note that the comparison of the final results obtained starting from the different initial approximations provides an estimation of the uncalculated higher-order contributions. The construction methods and application examples for the LDF and CH potentials can be found in Refs.~\cite{Sapirstein:2002:042501,pot:LDF,Sapirstein:2011:012504,Malyshev:2017:022512}.

In the $jj$ coupling, the unperturbed wave functions of the $^3P_{0}$ and $^3P_{2}$ levels under consideration are represented by the linear combinations of the Slater determinants of the Dirac wave functions with given values of the total angular momentum $J=0$ and $J=2$, namely, by the $1s^2\,(2s\,2p_{1/2})_0$ and $1s^2\,(2s\,2p_{3/2})_2$ states. By including the screening potential into the Dirac equation~(\ref{DirEq}) one can partly account for the effects of the interelectronic interaction. The remaining part of the electron-electron interaction as well as the interaction with the quantized electromagnetic field have to be considered by a perturbation theory. To construct the QED perturbation series we employ the two-time Green function (TTGF) method \cite{TTGF}. 

The numerical procedure which we use in the present work for the calculations of the binding energies of the $1s^2\,2s\,2p\, ^3P_{0,2}$ states is in general similar to the one described in details in Ref.~\cite{Malyshev:2017:022512}. It involves the rigorous calculations of the contributions corresponding to the one- and two-photon exchange Feynman diagrams and the evaluation of the one- and two-electron one-loop self-energy and vacuum-polarization corrections. All the calculations have been performed without any expansion in powers of the interaction with the binding potential in Eq.~(\ref{DirEq}). The many-electron QED contributions have been evaluated in Feynman and Coulomb gauges for the photons responsible for the electron-electron interaction. A good agreement between the calculations in both gauges is found out. In order to complete \textit{ab initio} treatment of the binding energies within the second order of the QED perturbation theory, one has to account for the contributions of the one-electron two-loop graphs. These corrections have been taken into account using the results presented in Refs.~\cite{Yerokhin:2015:033103,Yerokhin:2008:062510}. The third- and higher-order correlation effects are considered within the lowest-order relativistic (Breit) approximation. In the present work this contribution has been evaluated by the direct summation of the perturbation series in the framework of the recursive perturbation approach \cite{Glazov:2017:46,Malyshev:2017:022512}. Finally, one has to go beyond the so called external field approximation which treats the nucleus as a motionless source of the electrical field with an infinite mass and account for the contribution due to the nuclear recoil effect. In order to do so, we have extended the method applied in Ref.~\cite{Malyshev:2017:recoil:preprint} to the calculations of the energies of He-like ions to the case of Be-like ions. 

For comparison, we have also performed approximate (non-\textit{ab-initio} QED) relativistic calculations of all the possible levels of the $1s^2\,2s\,2p$ configuration. The calculations are based on the application of the large-scale CI-DFS method~\cite{Bratzev:1977:2655,Tupitsyn:2003:022511}, which treats the interelectronic interaction within the Breit approximation. In order to account for the QED and recoil effects, we include three corrections to the no-pair Dirac-Coulomb-Breit (DCB) Hamiltonian underlying the CI-DFS method. The first correction is the frequency-dependent ($\omega$-dependent) part of the Breit interaction in the Coulomb gauge. 
The nuclear recoil effect is calculated within the leading-order relativistic approximation and to all orders in $1/Z$. For this aim, the relativistic recoil operator~\cite{Shabaev:1985:394:note,Shabaev:1988:107:note} has been averaged with the many-electron CI-DFS wave functions, see Refs.~\cite{Tupitsyn:2003:022511,Kozhedub:2010:042513} for details. Finally, the radiative QED effects been estimated with the use of the model QED operator approach (QEDMOD)~\cite{Shabaev:2013:012513,Shabaev:2013:175}. This approach is much simpler and less accurate than the full-scale QED calculations, but it provides remarkably well results. 


%

         
\section{Numerical results and discussion \label{sec:result}}

In the present section we discuss our results obtained for the binding energies of Be-like ions. The individual contributions to the binding energies of the $1s^2\,2s\,2p$ electron configuration in berylliumlike xenon evaluated starting from the Coulomb nuclear potential and with the LDF and CH screening potentials are shown in Table~\ref{tab:3P0} for the $^3P_0$ level and in Table~\ref{tab:3P2} for $^3P_2$ level. In both tables the first line presents the zeroth-order value of the binding energy calculated using the one-electron Dirac energies from Eq.~(\ref{DirEq}). The Fermi model with a thickness parameter equal to 2.3~fm has been used to describe the nuclear charge distribution. The root-mean-square radii were taken from Ref.~\cite{Angeli:2013:69}. In the second row the first-order interelectronic-interaction contribution evaluated in the framework of the Breit approximation is given. The correction due to the energy dependence of the interelectronic-interaction operator is shown in the third line. The next two rows contain the second-order electron-electron interaction correction within the Breit approximation and the corresponding QED correction $E_{\rm int,QED}^{(2)}$. We note, that the second-order value $E_{\rm int,Breit}^{(2)}$ was obtained by calculating at zero-energy transfer in the Coulomb gauge with the negative-energy continuum contribution neglected. This approach to the Breit approximation differs, e.g., from the one used in Ref.~\cite{Yerokhin:2007:062501}, where the Breit part of the interelectronic-interaction operator was treated to the first order only and the negative-energy Dirac spectrum was partly taken into account. In the fifth line we present the third- and higher-order correlation effects evaluated within the Breit approximation by means of the recursive formulation of the perturbation theory. As in case of the second-order correction, the exchange by the Breit photons has been considered to all orders. The contributions of the first- and second-order one-loop self-energy and vacuum-polarization Feynman diagrams are collected in the next two rows. The last presented contribution within the external field approximation corresponds to the one-electron two-loop diagrams, it is labeled as $E_{\rm 2loop}^{(2)}$. The contributions due to the nuclear recoil effect calculated within the lowest-order relativistic approximation and the QED recoil effect are given in the rows $E_{\rm recoil,Breit}$ and $E_{\rm recoil,QED}$, respectively, see Ref.~\cite{Malyshev:2017:recoil:preprint} for details. Finally, the total values of the binding energies of the $1s^2\,2s\,2p\,^3P_0$ and $1s^2\,2s\,2p\,^3P_2$ states are presented in the last lines of Tables~\ref{tab:3P0} and \ref{tab:3P2}, respectively. One can see that the results of the calculations performed within the extended Furry picture with the LDF and CH screening potentials included into the unperturbed Hamiltonian are in good agreement with each other even despite of the different asymptotic behavior of the potentials employed. On the other hand, the value obtained starting from the Coulomb potential of the nucleus stands apart slightly. This results from the fact that the application of the extended Furry picture allows one to take into account the higher-order QED contributions partly. As the final theoretical values for both states we have chosen the values obtained for the LDF potential. The deviations from the Coulomb results were used in order to estimate the uncertainty associated with the uncalculated higher-order QED effects. 

Our final theoretical predictions for the binding energies of the $^3P_0$ and $^3P_2$ levels of the electron configuration $1s^2\,2s\,2p$ in berylliumlike xenon which have been obtained with the use of \textit{ab initio} method are given in Table~\ref{tab:binding}. The results of the approximate calculations are also given. The uncertainties indicated in the brackets are mainly due to the approximations employed in the calculations of the two-loop contributions \cite{Yerokhin:2015:033103,Yerokhin:2008:062510}, the uncertainty of the third- and higher-order interelectronic-interaction effects evaluation, and an estimation made for the uncalculated higher-order QED corrections. 


{
\renewcommand{\arraystretch}{1.15}
\begin{table}
\begin{center}
\caption{Individual contributions to the energy of the $1s^2\,2s\,2p \, ^3P_0$ state in berylliumlike xenon (in eV). Calculations by \textit{ab initio} method. See text for details. \label{tab:3P0}}
\end{center}
\begin{center}
\vspace*{-4mm}
\begin{tabular}{@{\quad}l@{\qquad}
                S[table-format=-7.5]
                S[table-format=-7.5]
                S[table-format=-7.5]
                }
                
\hline

Contribution   &  {Coulomb}  & {LDF} & {CH}  \\[1mm]
                
\hline                
                
$E_{\rm Dirac}^{(0)}$                &   -103574.0434  &   -97759.7240  &   -98287.2101  \\

$E_{\rm int,Breit}^{(1)}$            &      2650.2021  &    -3197.9357  &    -2669.1707  \\

$E_{\rm int,QED}^{(1)}$              &         0.0220  &        0.0168  &        0.0173  \\

$E_{\rm int,Breit}^{(2)}$            &       -37.6082  &       -3.7286  &       -5.0332  \\

$E_{\rm int,QED}^{(2)}$              &         0.2344  &        0.2544  &        0.2692  \\

$E_{\rm int,Breit}^{(\geqslant 3)}$  &         0.1207  &        0.0315  &        0.0407  \\

$E_{\rm SE+VP}^{(1)}$                &        94.4571  &       91.6963  &       91.2014  \\

$E_{\rm ScrSE+ScrVP}^{(2)}$          &        -2.2959  &        0.5099  &        1.0086  \\

$E_{\rm 2loop}^{(2)}$                &        -0.2492  &       -0.2492  &       -0.2492  \\

$E_{\rm recoil,Breit}$               &         0.3832  &        0.3833  &        0.3833  \\

$E_{\rm recoil,QED}$                 &         0.0415  &        0.0402  &        0.0400  \\[0.5mm]

\hline

$E_{\rm total}$                      &   -100868.7358  &  -100868.7050  &  -100868.7027  \\[1mm]

\hline

\end{tabular}%
\end{center}
\end{table}
}


{
\renewcommand{\arraystretch}{1.15}
\begin{table}
\begin{center}
\caption{Individual contributions to the energy of the $1s^2\,2s\,2p \, ^3P_2$ state in berylliumlike xenon (in eV). Calculations by \textit{ab initio} method. See the text for details. \label{tab:3P2}}
\end{center}
\begin{center}
\vspace*{-4mm}
\begin{tabular}{@{\quad}l@{\qquad}
                S[table-format=-7.5]
                S[table-format=-7.5]
                S[table-format=-7.5]
                }

\hline

Contribution   &  {Coulomb}  & {LDF} & {CH}  \\[1mm]
                
\hline   

$E_{\rm Dirac}^{(0)}$                &  -103147.3211  &   -97382.7538  &   -97909.8215  \\

$E_{\rm int,Breit}^{(1)}$            &     2584.8484  &    -3210.3295  &    -2681.9588  \\

$E_{\rm int,QED}^{(1)}$              &       -0.4225  &       -0.3700  &       -0.3698  \\

$E_{\rm int,Breit}^{(2)}$            &      -34.1690  &       -3.5831  &       -4.9175  \\

$E_{\rm int,QED}^{(2)}$              &        0.2546  &        0.2346  &        0.2529  \\

$E_{\rm int,Breit}^{(\geqslant 3)}$  &        0.0481  &        0.0292  &        0.0407  \\

$E_{\rm SE+VP}^{(1)}$                &       95.1540  &       92.3205  &       91.8276  \\

$E_{\rm ScrSE+ScrVP}^{(2)}$          &       -2.3474  &        0.5310  &        1.0275  \\

$E_{\rm 2loop}^{(2)}$                &       -0.2503  &       -0.2503  &       -0.2503  \\

$E_{\rm recoil,Breit}$               &        0.3807  &        0.3809  &        0.3809  \\

$E_{\rm recoil,QED}$                 &        0.0417  &        0.0404  &        0.0401  \\[0.5mm]

\hline

$E_{\rm total}$                      &  -100503.7827  &  -100503.7502  &  -100503.7481  \\[1mm]

\hline

\end{tabular}%
\end{center}
\end{table}
}


{
\renewcommand{\arraystretch}{1.15}
\begin{table}
\begin{center}
\caption{Binding energies of the $^3P_0$ and $^3P_2$ levels of the $1s^2\,2s\,2p$ electron configuration in berylliumlike xenon (in eV). \label{tab:binding}}
\end{center}
\begin{center}
\vspace*{-4mm}
\begin{tabular}{@{\quad}
                c@{\qquad}                
                S[table-format=-7.3(2)]@{\qquad}
                l@{\quad}
                }
\hline

 Level & {Binding energy} & Reference \\
 
\hline
 
 $^3P_0$      &   -100868.705(65)   &   This work, \textit{ab initio}         \\
 
              &   -100868.888       &   This work, approximate                \\
 
              &   -100869.08        &   Gu \textit{et al.} \cite{Gu:2005:267} \\
                 
\hline                 
 
 $^3P_2$      &   -100503.750(65)   &   This work, \textit{ab initio}        \\
 
              &   -100503.882       &   This work, approximate                \\
  
              &   -100503.74        &   Gu \textit{et al.} \cite{Gu:2005:267} \\ 

\hline

\end{tabular}%
\end{center}
\end{table}
}


{
\renewcommand{\arraystretch}{1.15}
\begin{table}
\begin{center}
\caption{Transition energy from the $1s^2\,2s\,2p\,^3P_2$ state to the $1s^2\,2s^2\,^1S_0$ ground state in berylliumlike xenon (in eV). \label{tab:excitation}}
\end{center}
\begin{center}
\vspace*{-4mm}
\begin{tabular}{@{\quad}
                S[table-format=3.3(2)]@{\qquad}                
                l@{\qquad}
                l@{\quad}
                }
\hline

 {Transition energy} & Reference & Work \\
 
\hline

 469.230(90)   &   This work, \textit{ab initio}                          &  \multirow{6}{*}{Theory}  \\
 
 469.572       &   This work, approximate                                 &                           \\ 
 
 469.449       &   Cheng \textit{et al.} \cite{Cheng:2008:052504}         &                           \\
 
 470.004       &   Gu \textit{et al.} \cite{Gu:2005:267}         		  &                           \\
  
 469.25        &   Safronova \cite{Safronova:2000:1213}                   &                           \\  
 
 469.386       &   Safronova \textit{et al.} \cite{Safronova:1996:4036}   &                           \\
  
\hline

 469.474(81)   &   Bernhardt \textit{et al.} \cite{Bernhardt:2015:144008} &  Experiment               \\

\hline

\end{tabular}%
\end{center}
\end{table}
}


{
\renewcommand{\arraystretch}{1.15}
\begin{table}
\begin{center}
\caption{Transition energy from the $1s^2\,2s\,2p\,^3P_2$ state to the $1s^2\,2s\,2p\,^3P_0$ state in berylliumlike xenon (in eV). \label{tab:PP_transition}}
\end{center}
\begin{center}
\vspace*{-4mm}
\begin{tabular}{@{\quad}
                S[table-format=3.3(2)]@{\qquad}                
                l@{\quad}
                }
\hline

 {Transition energy} & Reference  \\
 
\hline

 364.955(45)   &   This work, \textit{ab initio}                           \\ 
 
 365.006       &   This work, approximate                                  \\ 
 
 364.974       &   Cheng \textit{et al.} \cite{Cheng:2008:052504}          \\
 
 365.341       &   Gu \textit{et al.} \cite{Gu:2005:267}         		   \\
 
 364.904       &   Safronova \textit{et al.} \cite{Safronova:1996:4036}    \\
 
\hline

\end{tabular}%
\end{center}
\end{table}
}

\begin{table}
\begin{center}
\caption{\label{tab:Ne_appr} 
            Individual contributions to the energy levels of the $1s^2\,2s\,2p$ electron configuration in  berylliumlike neon (in eV). Calculations by the approximate method. 
            See text for details.}

\end{center}      

\begin{center}
\begin{tabular}{@{\quad}
                 l@{\qquad}
                 S[table-format=-7.4,group-separator=]@{\qquad}
                 S[table-format=-7.4,group-separator=]@{\qquad}
                 S[table-format=-7.4,group-separator=]@{\qquad}
                 S[table-format=-7.4,group-separator=]
                                  @{\quad}}
\hline
 \multicolumn{1}{c}{ Contribution } & 
 \multicolumn{1}{c}{$ ^3P_0 $} & \multicolumn{1}{c}{$ ^3P_1 $} &
 \multicolumn{1}{c}{$ ^1P_1 $} & \multicolumn{1}{c}{$ ^3P_2 $} 
\\
\hline
  $ E_{{\rm int,Breit}} $    &    -2990.945 &  -2990.889   &   -2978.086  &   -2990.766                \\ 
  $ E_{{\rm int,Breit\textnormal{-}fr}} $   &          -0.000 &        -0.000   &         -0.000  &         -0.000                \\ 
  $ E_{{\rm QEDMOD}} $   &           0.286 &         0.287   &          0.286  &          0.287                \\ 
  $ E_{{\rm recoil,Breit}} $   &           0.080 &         0.080   &          0.079  &          0.080                \\ 
  $ E_{{\rm total}} $    &       -2990.579 &     -2990.523   &      -2978.293  &      -2990.399                \\ 
\hline

\end{tabular}
\end{center}
\end{table}
\begin{table}
\begin{center}
\caption{\label{tab:Fe_appr} 
            Individual contributions to the energy levels of the $1s^2\,2s\,2p$ electron configuration in  berylliumlike iron (in eV). Calculations by the approximate method. 
            See text for details.}

\end{center}      

\begin{center}
\begin{tabular}{@{\quad}
                 l@{\qquad}
                 S[table-format=-7.4,group-separator=]@{\qquad}
                 S[table-format=-7.4,group-separator=]@{\qquad}
                 S[table-format=-7.4,group-separator=]@{\qquad}
                 S[table-format=-7.4,group-separator=]
                                  @{\quad}}
\hline
 \multicolumn{1}{c}{ Contribution } & 
 \multicolumn{1}{c}{$ ^3P_0 $} & \multicolumn{1}{c}{$ ^3P_1 $} &
 \multicolumn{1}{c}{$ ^1P_1 $} & \multicolumn{1}{c}{$ ^3P_2 $} 
\\
\hline
  $ E_{{\rm int,Breit}} $    &   -22068.071 &  -22064.252   &  -22017.978  &  -22052.787                \\ 
  $ E_{{\rm int,Breit\textnormal{-}fr}} $   &           0.000 &        -0.002   &         -0.007  &         -0.008                \\ 
  $ E_{{\rm QEDMOD}} $   &           8.021 &         8.028   &          8.037  &          8.059                \\ 
  $ E_{{\rm recoil,Breit}} $   &           0.204 &         0.204   &          0.203  &          0.204                \\ 
  $ E_{{\rm total}} $    &      -22059.846 &    -22056.022   &     -22025.819  &     -22044.533                \\ 
\hline

\end{tabular}
\end{center}
\end{table}
\begin{table}
\begin{center}
\caption{\label{tab:Xe_appr} 
            Individual contributions to the energy levels of the $1s^2\,2s\,2p$ electron configuration in berylliumlike xenon (in eV). Calculations by the approximate method. 
            See text for details.}

\end{center}      

\begin{center}
\begin{tabular}{@{\quad}
                 l@{\qquad}
                 S[table-format=-7.4,group-separator=]@{\qquad}
                 S[table-format=-7.4,group-separator=]@{\qquad}
                 S[table-format=-7.4,group-separator=]@{\qquad}
                 S[table-format=-7.4,group-separator=]
                                  @{\quad}}
\hline
 \multicolumn{1}{c}{ Contribution } & 
 {$ ^3P_0 $} & {$ ^3P_1 $} &
 {$ ^1P_1 $} & {$ ^3P_2 $} 
\\
\hline
  $ E_{{\rm int,Breit}} $    &  -100961.330 &  -100938.536   &  -100533.223  &  -100596.618                \\ 
  $ E_{\mathrm{int,Breit\textnormal{-}fr}} $   &           0.016 &         0.012   &         -0.367  &         -0.371                \\ 
  $ E_{{\rm QEDMOD}} $   &          92.042 &        92.030   &         92.654  &         92.726                \\ 
  $ E_{{\rm recoil,Breit}} $   &           0.383 &         0.383   &          0.380  &          0.380                \\ 
  $ E_{{\rm total}} $    &     -100868.888 &   -100846.111   &    -100625.864  &    -100503.882                \\ 
\hline

\end{tabular}
\end{center}
\end{table}

In order to perform a more comprehensive analysis of the results obtained by \textit{ab initio} method, in Table~\ref{tab:excitation} we present the transition energy from the $1s^2\,2s\,2p\,^3P_2$ state to the $1s^2\,2s^2\,^1S_0$ ground one for Be-like xenon. The result evaluated with the use of the alternative approximate approach is shown as well. The transition energy was calculated by subtracting the binding energy of the ground state from the binding energy of the state under consideration. In order to do so, we have recalculated the ground-state binding energy with \textit{ab initio} method employed in the present work instead of using the value presented in Ref.~\cite{Malyshev:2014:062517}. There are two main reasons for it. First, in Ref.~\cite{Malyshev:2014:062517} the third- and higher-order interelectronic-interaction contributions have been evaluated with the use of the CI-DFS method in contrast to the present work where the recursive perturbation theory has been employed. The corresponding contribution has changed within the designated error bar. Second, the values of the two-loop corrections used now and then also differ slightly. The deviation of this correction lies within the corresponding error bar too. Thus, in order to obtain the transition energy from the $1s^2\,2s\,2p\,^3P_2$ state to the ground one we have used the new theoretical value for the energy of the ground state, $-100\,972.981(85)$~eV, instead of the old value, $-100\,972.921(85)$~eV. The uncertainty of this transition energy is determined mainly by the uncertainty of the ground-state energy. The closeness of the states $1s^2\,2s^2$ and $1s^2\,(2p_{1/2})^2$ with the same symmetry makes the convergence of the perturbation series for the ground state slow. 

In Table~\ref{tab:excitation}, we compare our theoretical prediction for the $2s\,2p\,^3P_2 - 2s^2\,^1S_0$ transition energy with the results of the previous relativistic calculations and the high-precision measurement. The discrepancy with the experimental value by Bernhardt \textit{et al.} \cite{Bernhardt:2015:144008} is observed. The reason of this discrepancy is unclear to us now. The approximate value is surprisingly closer to the experiment than our \textit{ab initio} result. It is likely an accidental coincidence, since the uncertainty of the approximate approach is significantly larger than the uncertainty of \textit{ab initio} method. As it was noted above, numerous theoretical results for the $2s\,2p\,^3P_2 - 2s^2\,^1S_0$ transition energy demonstrate the significant scatter. The best agreement with our value is found for the calculations performed by U.~Safronova~\cite{Safronova:2000:1213}. We note that the result of the previous evaluation accomplished by the same authors lies further from our one \cite{Safronova:1996:4036}. Nevertheless, it is expected that our theoretical predictions obtained within \textit{ab initio} method must have higher accuracy compared to the previous calculations, since we have evaluated the many-electron QED corrections rigorously without the application of any one-electron or semiempirical approximation. 

In Table~\ref{tab:PP_transition} we compare the theoretical energies of the transition from the $1s^2\,2s\,2p\,^3P_2$ state to the $1s^2\,2s\,2p\,^3P_0$ state in berylliumlike xenon. This transition energy does not depend on the ground-state binding energy. As a result, the perturbation theory converges better, and this transition can be studied to a higher accuracy. Indeed, from Table~\ref{tab:PP_transition} one can see that this is the case. The scatter of the results obtained in the framework of the different calculations is much smaller than for the $2s\,2p\,^3P_2 - 2s^2\,^1S_0$ transition energy in Table~\ref{tab:excitation}.

Let us now discuss in more details the results obtained with the use of the approximate method.
The individual contributions to the binding energies of the $1s^2\,2s\,2p$ electron configuration in berylliumlike neon, iron, and xenon are shown in Tables~\ref{tab:Ne_appr}, \ref{tab:Fe_appr}, and \ref{tab:Xe_appr}, respectively. In all tables the first line presents the Dirac-Coulomb-Breit energy $ E_{{\rm int,Breit}} $ evaluated by means of the CI-DFS method. All presented digits are relevant within the approach under consideration. In the second row the frequency-dependent Breit correction $E_{{\rm int,Breit\textnormal{-}fr}}$ is given. 
We note that in the present work, on the contrary to the approach sometimes used in the literature, see, e.g., Ref.~\cite{Chen:1995:266}, we do not construct the whole CI matrix with the frequency-dependent Breit interaction included in order to obtain this correction. Instead, we apply the frequency-dependent Breit interaction for the reference-configuration-state functions only, as it was suggested in Ref.~\cite{Yerokhin:2012:042507}. The correction due to the radiative QED effects $E_{{\rm QEDMOD}}$ which was obtained within the model QED operator approach is shown in the third line. The uncertainty associated with the approximate treatment of the Lamb shift can be estimated properly only after the thorough comparison with the rigorous calculations. Based on the comparison of the approximate results with the results of \textit{ab initio} QED calculations for berylliumlike xenon and data from Refs.~\cite{Yerokhin:2014:022509,Yerokhin:2015:054003} for berylliumlike argon and iron we can conclude that uncertainty of the model QED operator approach for the low-lying states of middle- and low-$Z$ berylliumlike ions does not exceed $2\%$. The contribution due to the nuclear recoil effect evaluated within the leading-order relativistic approximation and to all orders in $1/Z$ is given in the row labeled $E_{\rm recoil,Breit}$. Finally, the total values of the binding energies of the low-lying singly excited stated in beryllilike neon, iron, and xenon are presented in the last lines of the Tables~\ref{tab:Ne_appr}, \ref{tab:Fe_appr}, and \ref{tab:Xe_appr}, respectively. One can see that the results of the calculations performed by approximate method are in reasonable agreement with \textit{ab initio} QED calculations performed for Be-like xenon. We also note that the approximate approach allows one to carry out calculations for (quasi)degenerate states in the same manner as for single states. 



         
\section{Summary}

To summarize, in this paper we have evaluated the binding energies of the $^3P_0$ and $^3P_2$ levels of the $1s^2\,2s\,2p$ electron configuration in berylliumlike xenon. Employed \textit{ab initio} method combines the rigorous treatment within the first and second orders of the QED perturbation theory with the third- and higher-order interelectronic-interaction contributions calculated in the framework of the lowest-order relativistic approximation. The QED nuclear recoil effect has been evaluated using the independent electron approximation. The approximate method which is based on the configuration-interaction Dirac-Fock-Sturm method merged with the model QED approach has been applied to the calculations of the energy levels of the selected berylliumlike ions as well. The obtained theoretical predictions are compared with the results of the previous relativistic calculations and experiment. The discrepancy between the result of our \textit{ab initio} calculation and the experimental value is found out. Further work both from theoretical and experimental sides is urgent in order to clarify the reasons of this discrepancy. 

         
\section*{Acknowledgments}

This work was supported by RFBR (Grant No.~16-02-00334), by SPSU-DFG (Grants No.~11.65.41.2017 and No.~STO~346/5-1), and by SPSU (Grant No.~11.40.538.2017).



%

%

\end{document}